\documentclass[12pt]{article}

\begin{document}

\begin{center}
{\large {\bf An exact self-similar solution for an expanding  ball of 
radiation}}
\end{center}

\vskip 0.8cm

\begin{center}
{ J. Ponce de Leon}\footnote{Correspondence to:
J. Ponce de Leon. E-mail:jponce@upracd.upr.clu.edu}
\vskip 0.2cm
{\it Department of Physics,
University of Puerto Rico, P.O. Box 23343, San Juan, PR 00931, USA.\\}
\vskip 0.6cm

{P.S. Wesson}
\vskip 0.2cm
{\it Department of Physics, University of Waterloo, Waterloo, Ontario N2L 
3G1, Canada.\\}
\vskip 0.6cm
{(November 2005)}

\end{center}

\begin{abstract}
We give an exact solution of the $5D$ Einstein equations which in $4D$ can 
be interpreted as
a spherically symmetric dissipative distribution of matter, with heat flux, whose effective density and 
pressure are nonstatic, nonuniform, and satisfy the equation of state
of radiation. The matter satisfies the usual energy and thermodynamic conditions. The energy density and temperature are related by the Stefan-Boltzmann law. The solution admits a homothetic Killing vector in $5D$, which 
induces the existence of self-similar symmetry in $4D$, where the line 
element as well as the  dimensionless matter quantities are invariant under 
a simple ``scaling" group.

\end{abstract}

\medskip



\newpage

\section{Introduction}
In recent years, there has been an explosion of interest in solutions of the
five-dimensional field equations of general relativity.  This is due mainly
to two things. Physically, it is now generally acknowledged that the
unification of gravity with the interactions of particle physics is best
approached through the extension of spacetime to higher dimensions, of which
(noncompactified) Kaluza-Klein space with five dimensions is the basic
example. Mathematically, it is now accepted that Campbell's theorem
guarantees that any solution of the four-dimensional Einstein field
equations with matter can be derived from a solution of the five-dimensional
Kaluza-Klein equations in apparent vacuum, provided an appropriate
identification is made for the energy-momentum tensor \cite{Campbell}-\cite{Seahra}. 

In the present
paper, we wish to focus on one situation which is fundamental to
astrophysics but has not received much attention.  Namely, that
of a non-static ball of radiation-like matter which is spherically symmetric
in ordinary $3D$ space, and whose properties are described by an exact
solution which has  self-similar symmetry.

The latter symmetry has been extensively applied in studies of fluid flows
in accordance with the equations of Newton and Einstein \cite{Sedov}-\cite{Cahill}. 
 It's usefulness
lies in the fact that when the equations are algebraically complicated, it
is sometimes possible to find solutions which depend on a {\it ratio}
of the independent variables (rather than on these separately), and that
such solutions are frequently realized to a good approximation in the real
world.  This correspondence arises, for example, when a solution does not
contain any arbitrary length scales, because the physical situation it
describes does not contain a sharp boundary \cite{Henriksen}-\cite{Jpdel5}.

While the self-similar symmetry has been extensively applied to obtain exact
solutions of the 4D Einstein equations of general relativity, it has not
been much studied in the context of the 5D Kaluza-Klein equations.  These
are currently the topic of much research, under the names of induced-matter
and membrane theory.  In the former approach, the fifth dimension is
responsible for the matter/energy content of the universe \cite{Wessonbook}; whereas in the
latter approach, the fifth dimension provides a singular surface around
which are concentrated the particle interactions of the universe \cite{Maartens}.  It has
recently become clear that the field equations of these two approaches are
basically the same \cite{Jpdel3}, so in the present paper we will adopt a generic
approach, and elucidate the physics of a self-similar exact solution of the
5D field equations.

These field equations are normally taken to be those obtained by setting the
higher-dimensional Ricci tensor to zero.  In this regard, they are the
extensions of the Einstein equations of general relativity, as verified by
the classic solar-system tests of that theory.  However, the 5D equations
are richer than the 4D ones, which means that Birkhoff's theorem does not
carry over in simple form to 5D from 4D \cite{Billyard}.  Nevertheless, the 5D case is
known to agree with the classic and other tests of astrophysics, and also
gives models which are in agreement with data from cosmology \cite{JpdelFRW}.  This
concordance, along with the promise of new physics, underlies the current
upsurge in interest in exact solutions of the 5D field equations.

However, a gap in our knowledge exists as regards the subject of this paper,
namely the case of a non-static distribution of matter which is spherically
symmetric in 3D space and has an equation of state of the hot or radiation
type.  Such a solution can be expected to be relevant to the early
universe, especially in regard to the growth of perturbations in the early
stages of inflation and to the related topic of galaxy formation.

Other solutions of this type exist in the literature \cite{Wessonbook}.  They are usually derived by solving the 5D field equation $R_{AB}=0$,
and then using the standard technique in which the solution is re-expressed
as one with $G_{\alpha \beta } = T_{\alpha \beta }$, where the last object
describes the {\it effective} properties of matter \cite{JpdelEMT} in accordance with Campbell's
theorem (here $R_{AB}$ is the 5D Ricci tensor, and $G_{\alpha
\beta }$, $T_{\alpha \beta }$ are the 4D Einstein and energy-momentum
tensors).  The details of this procedure are by now well known to
researchers in the field, and algebraically tedious. 

In the present paper we give a solution of the 
$5D$ equations which in $4D$ represents an expanding spherical distribution  
of matter whose effective equation of state is that of radiation.
The solution may be verified by a fast
software package such as GRTensor \cite{Lake}, and the long form of the field equations
may be found in the literature \cite{Wessonbook}, \cite{JpdelWesson}.  

Our solution
has some unique properties: (i) it is nonstatic and spatially nonuniform;
(ii) it is curved in 5D, but (iii) admits a homothetic killing vector, and
(iv) exists for both signs of the extra part of the metric. \ In these
regards, it is different and more realistic than other solutions which have
appeared, and may in the future deserve more intensive study.

\section{A solution for an expanding ball of radiation}

For an early universe, we are interested in solutions that describe 
expanding fluids  with an equation of state which is radiation-like. These 
should be spherically symmetric in ordinary $3D$ space. Therefore, we choose 
the line element as
\begin{equation}
\label{metric}
{d{\cal{S}}^2} = e^{\nu}dt^2 - e^{\lambda}dr^2 - R^2(d\theta^2 + sin^2 
\theta d\phi^2) + \epsilon \Phi^2 dy^2,
\end{equation}
where $y$ is the coordinate along the extra dimension and $t$, $r$, $\theta$ 
and $\phi$ are the usual coordinates for a spacetime with spherically 
symmetric spatial sections. The factor $\epsilon$ in front of $\Phi^2$ can 
be either $- 1$ or $+ 1$ depending on whether the extra dimension is 
spacelike or timelike, respectively. In what follows we choose units such 
that the speed of light and the gravitational constant are $c = 1$ and $8 
\pi G = 1.$

In (\ref{metric}) the metric coefficients $\nu$, $\lambda$, $R$ and $\Phi$ 
are, in principle, functions of the extra coordinate. However, we assume 
from the beginning that $\nu(r,t)$, $\lambda(r,t)$, $R(r,t)$ and $\Phi(r,t)$ 
are not functions of $y$. In this way  we assure  the radiation-like nature 
of the effective matter in $4D$

The metric in $5D$ is a solution of the field equations \cite{Wessonbook}, 
which in terms of the Ricci tensor read $R_{AB} = 0$ $(A, B = 0, 1, 2, 3, 
4)$. Here, for brevity we omit the details of the derivation and state the 
solution
\begin{equation}
\label{the solution}
{d{\cal{S}}^2} = B^2 \left[\frac{3r^2}{\alpha^2}dt^2 - t^2 dr^2 - \frac{t^2 
r^2}{(3 - \alpha^2)}(d\theta^2 + sin^2 \theta d\phi^2)\right] \pm C^2 
r^{2(\alpha + 1)}t^{2(\alpha + 3)/\alpha}dy^2,
\end{equation}
which  may be verified by computer. Here $\alpha$ is a dimensionless 
parameter in the range $0 < \alpha^2 < 3$, whereas $B$ and $C$ are arbitrary 
constants with the dimensions $L^{- 1}$ and $L^{- (\alpha^2 + 2\alpha + 
3)/\alpha}$, respectively. The parameter $\alpha$ is related to the time 
evolution of the extra dimension. Indeed, from (\ref{the solution}) it 
follows that
\begin{equation}
\label{change in Phi}
\frac{1}{\Phi}\frac{d\Phi}{dt} = \frac{(\alpha + 3)}{\alpha t}.
\end{equation}
Since $0 < \alpha^2 < 3$, it follows that  either $- \sqrt{3} < \alpha < 0$ 
or $0 < \alpha < \sqrt{3}$. In both cases $(\alpha + 3)$ is positive. 
Therefore, for a negative (positive) $\alpha$ the size of the extra 
dimension decreases (increases) monotonically with time.

For the four-dimensional interpretation of the solution we identify our 
spacetime with a  hypersurface, located at some value of $y$, orthogonal 
to the extra dimension. This is the common assumption in STM and brane 
theory. Despite the differences between them, the effective matter content 
is the same in both approaches \cite{Jpdel3}. The appropriate formulae for 
the calculation of the energy-momentum-tensor $T_{\mu\nu}$ for the metric 
(\ref{metric}) can be found in Ref. $[13]$. Using (\ref{the solution}) we 
get
\begin{eqnarray}
\label{matter distribution}
T_{0}^{0} &=& \frac{2}{B^2 t^2 r^2},\nonumber \\
T^{1}_{1} &=& - \frac{2 \alpha^2}{3 B^2 t^2 r^2},\nonumber \\
T_{2}^{2} &=&  - \frac{(3 - \alpha^2)}{3 B^2 t^2 r^2} = T_{3}^{3},\nonumber 
\\
T_{0}^{1} &=& - \frac{2}{B^2 t^3 r}.
\end{eqnarray}
We emphasize that Campbell's theorem guarantees the validity of Einstein's 
equations $G_{\mu\nu} = T_{\mu\nu}$. For the physical interpretation of the 
spherical distribution of matter (\ref{matter distribution}) we notice that: 
(i) for  $\alpha^2 \neq 1$ the principal stresses are unequal, (ii) there is 
a radial flux of energy $T_{01} \neq 0$, (iii) it satisfies the energy 
conditions
\begin{equation}
T_{0}^{0} > 0, \;\;\; (\alpha^2 < 3),\;\;\; T_{0}^{0} > |T_{i}^{i}|, \;\;\; 
(i = 1, 2, 3),
\end{equation}
and (iv) the trace of the energy-momentum-tensor vanishes, viz.,
\begin{equation}
\label{equation of state}
T_{0}^{0} + T_{1}^{1} + T_{2}^{2} + T_{3}^{3} = 0.
\end{equation}
This is the usual relation  for ultra-relativistic particles or photons,  
and in terms of the average pressure $\bar{p} = - (T_{1}^{1} + T_{2}^{2} + 
T_{3}^{3})/3$ and density $\rho = T_{0}^{0}$ is commonly written as $\rho = 
3\bar{p}$.

\section{Model for the matter}

In order to ascribe a precise meaning to the components (\ref{matter 
distribution}) it is necessary to adopt a model for the matter. The general 
form of the energy-momentum tensor for anisotropic matter with energy flux 
can be covariantly given as
\begin{equation}
\label{general EMT}
T_{\mu\nu} = (\rho + p_{\perp})u_{\mu}u_{\nu} - p_{\perp}g_{\mu\nu} +(p - 
p_{\perp})\chi_{\mu}\chi_{\nu}  + q_{\mu}u_{\nu} + q_{\nu}u_{\mu},
\end{equation}
where $u^{\mu}$ is the four-velocity, $\chi^{\mu}$ is a unit spacelike 
vector orthogonal to  $u^{\mu}$, $\rho$ is the energy density,  $p$ is the 
pressure in the direction of $\chi_{\mu}$, and $p_{\perp}$ is the pressure 
in the two-space orthogonal to $\chi_{\mu}$. The energy flux is described by 
the $4$-vector $q_{\mu}$ and obeys $q_{\mu}u^{\mu} = 0$.

We now proceed to discuss the matter quantities entering (\ref{general EMT}) 
in the comoving and non-comoving frames.
\subsection{Comoving frame}
For a spherical distribution of matter, in a comoving frame we have
\begin{equation}
u_{\mu} = \delta^{0}_{\mu}e^{\nu/2}, \;\;\; \chi_{\mu} = \delta^{1}_{\mu}e^{ 
  \lambda/2}, \;\;\; q_{\mu} = \delta^{1}_{\mu} e ^{ \lambda/2}q ,
\end{equation}
where $q = (- q_{\mu}q^{\mu})^{1/2}$. Consequently, for the metric (\ref{the 
solution}) the matter quantities are as follows
\begin{eqnarray}
\label{model in com. system}
\rho = \frac{2}{B^2 t^2 r^2},\;\;\;
p = \frac{2\alpha^2}{3 B^2t^2 r^2},\;\;\;
p_{\perp} = \frac{(3 - \alpha^2)}{3 B^2 t^2 r^2},
\end{eqnarray}
and
\begin{equation}
\label{heat in com. system}
q =  \frac{2 \alpha}{\sqrt{3} B^2 t^2 r^2}.
\end{equation}
We see that for $\alpha^2 = 1$, the radial and tangential stresses are 
identical to each other $p = p_{\perp}$, i.e., it is a perfect fluid with an 
equation of state $\rho = 3 p$. For any other $\alpha^2 \neq 1$ in the range 
$\alpha^2 < 3$ the density and principal pressures are positive, $\rho = p + 
2 p_{\perp}$. Besides, $\rho > |q|$, which is physically reasonable. 
In the present case, 
  the expansion $\Theta = u^{\mu}_{;\mu}$ is given by
\begin{equation}
\Theta = \frac{\alpha \sqrt{3}}{B t r}.
\end{equation}
We note that $(\alpha/B)$ is a positive quantity because the four velocity is future-oriented, i.e., $u_{0}$ and $u^{0}$ are chosen to be positive. Consequently, the fluid under consideration is in continuous expansion.

\subsection{Non-comoving frame}
Thus, in the comoving frame the difference $|\alpha^2 - 1|$ can be used to 
measure the ``degree" of anisotropy of the fluid. We now proceed to show 
that for any value of $\alpha$, in the allowed range, one can always ``jump" 
to a moving frame where the fluid shows isotropic pressure.

For a spherical distribution of matter, in a radially moving frame the 
four-velocity and energy flux can be written as

\begin{eqnarray}
\label{non-comoving frame}
u^{\mu} &=& (e^{ - \nu/2}\cosh\omega,\;\;e^{- \lambda/2}\sinh \omega, 
\;\;0,\;\;0),\nonumber \\
q_{\mu} &=& q \;(- e^{\nu/2}\sinh\omega,\;\;e^{\lambda/2}\cosh \omega, 
\;\;0,\;\;0),
\end{eqnarray}
where $\omega$ is a parameter that measures the radial velocity $v$, viz.,
\begin{equation}
\cosh\omega = \frac{1}{\sqrt{1 - v^2}}, \;\;\;\sinh\omega = \frac{v}{\sqrt{1 
- v^2}}.
\end{equation}
The introduction of the hyperbolic functions assures the fulfillment of the 
conditions $u_{\mu}u^{\mu} = 1$ and $u_{\mu}q^{\mu} = 0$.

For perfect fluid $(p = p_{\perp})$, the energy-momentum tensor 
(\ref{general EMT}) with components (\ref{matter distribution}) in the frame 
(\ref{non-comoving frame}) yields
\begin{eqnarray}
\label{density and pressure in non-comoving frame}
\rho &=& (T_{0}^{0} + T_{1}^{1} - T_{2}^{2}) = \frac{(3 - \alpha^2)}{B^2 t^2 
r^2},\nonumber \\
p &=& - T_{2}^2 = \;\;\frac{\rho}{3}
\end{eqnarray}
and
\begin{equation}
q = \zeta \rho,
\end{equation}
where for simplicity we have introduced the dimensionless quantity $\zeta$.
The explicit expressions for $T_{0}^{0}$, $T_{1}^{1}$ and $T^{1}_{0}$ from 
(\ref{general EMT}) are
\begin{eqnarray}
4\cosh^2\omega - 6 \zeta \sinh\omega \cosh\omega &=& \frac{(9 - \alpha^2)}{(3 
- \alpha^2)},\nonumber \\
4 \sinh^2\omega - 6\zeta\sinh\omega \cosh\omega &=& \frac{3(\alpha^2 - 1)}{(3 
- \alpha^2)}, \nonumber \\
3\zeta(\cosh^2\omega + \sinh^2\omega ) - 4 \sinh\omega \cosh\omega &=& 
\frac{2 \sqrt{3}\alpha}{(3 - \alpha^2)},
\end{eqnarray}
respectively. These equations can be regarded as a {\it system} of equations in the regions $ 0 < \alpha \leq 1$ and $- \sqrt{3} < \alpha \leq - 1$. The solution is 
\begin{equation}
\zeta = \frac{|\alpha|}{\alpha \sqrt{3}}, \;\;\cosh\omega = \sqrt{\frac{9 + 
\alpha^2 - 6|\alpha|}{2(3 - \alpha^2)}}, \;\; \sinh\omega = \sqrt{\frac{3 + 
3\alpha^2 - 6|\alpha|}{2(3 - \alpha^2)}}.
\end{equation}
Consequently, we find
\begin{equation}
\label{radial velocity}
q = \frac{|\alpha|(3 - \alpha^2)}{\alpha \sqrt{3}B^2 t^2 r^2}, \;\;\;\;\;v = 
\sqrt{\frac{3 + 3\alpha^2 - 6|\alpha|}{9 + \alpha^2 - 6|\alpha|}}.
\end{equation}
Thus, for $\alpha = \pm 1$ the radial velocity vanishes and we recover the 
expressions (\ref{model in com. system}) and (\ref{heat in com. system}) for 
perfect fluid in the comoving frame, as expected. It should be noted that 
the matter quantities are not affected by the signature of the extra 
dimension or the sign of $\alpha$. The latter affects the evolution of the 
extra dimension (\ref{change in Phi}).
\subsection{Thermodynamic relations}

In order to ensure that a model with heat conduction is physically acceptable, besides the usual energy conditions, an appropriate set of thermodynamic relations must be satisfied. These are: (i) baryon conservation

\begin{equation}
\label{conservation of n}
(nu^{\mu})_{;\mu} = 0,
\end{equation}
where $n$ is the particle density; (ii) the first law of thermodynamics 
\begin{equation}
\label{first law of thermodynamics}
T d(S/n) = d (\rho/n) + p d(1/n),
\end{equation}
where $T$ is the temperature and $S$ is the entropy density; (iii) the temperature gradient law
\begin{equation}
\label{Gibbs relation}
q_{\mu} = - k [(\delta_{\mu}^{\nu} - u_{\mu}n^{\nu})T_{,\nu} + T u_{\mu;\nu}u^{\nu}],
\end{equation}
where $k$ is the thermal conductivity which must satisfy
\begin{equation}
\label{condition on k}
k \geq 0,
\end{equation}
and (iv) positive entropy production
\begin{equation}
\label{positive entropy production}
S^{\mu}_{;\mu} \geq  0,
\end{equation}
where $S^{\mu} = S u^{\mu} + q^{\mu}/T$.

The model under consideration here satisfies all the above conditions. Indeed, integrating  (\ref{conservation of n}), with $u^{\mu}$ from (\ref{non-comoving frame}), we obtain 
\begin{equation}
\label{n for the model here}
n = \frac{n_{0}}{(B^2 t r)^3},
\end{equation}
where $n_{0}$ is a constant. We see that the number of particles decreases with time, at any given location,  which is consistent with the expansion of the fluid.

The temperature $T$ occurs  in (\ref{first law of thermodynamics})
as an integrating factor. In our case, for    (\ref{density and pressure in non-comoving frame}) and (\ref{n for the model here}) we find this factor to be $T \sim  (rt)^{\beta}$, where $\beta $ is some constant. If we take $\beta = - 1/2$, then the density and temperature are related by the Stefan-Boltzmann law. Namely,   
\begin{equation}
\label{Stefan-Bolzmann law}
\rho = B^2 (3 - \alpha^2)\left(\frac{T}{T_{0}}\right)^4,
\end{equation}
with  
\begin{equation}
T = \frac{T_{0}}{B \sqrt{rt}},
\end{equation}
where $T_{0}$ is a constant with the appropriate units. We note that $(T_{0}/B) $ must be positive in order to ensure $T > 0$. In addition, in view of the fact that the temperature is a decreasing function of the radial coordinate, the heat flux is directed outward. 

For the thermal conductivity $k$ we find
\begin{equation}
k = \frac{2 (3 - \alpha^2)\sqrt{1 - v^2}}{3 |\alpha| \sqrt{3 rt}(1 + \alpha v/ \sqrt{3})}\left(\frac{\alpha}{B}\right)\left(\frac{B}{T_{0}}\right).
\end{equation}
Since $(B/T_{0})$ and $(\alpha/B)$ are positive quantities, it follows that $k$ is positive in the allowed range of $\alpha$, which is consistent with (\ref{condition on k}).

At this point we should mention that condition  (\ref{positive entropy production}) is ensured by conditions (\ref{Gibbs relation}) and (\ref{condition on k}) so it is not a separate condition. Therefore, the conclusion from the above discussion is that our model satisfies appropriate physical requirements in the regions $ 0 < \alpha \leq 1$ and $- \sqrt{3} < \alpha \leq - 1$. 
 
\section{Properties of the solution}
The solution under consideration has a number of interesting mathematical 
properties.

Firstly, the line element (\ref{the solution}) admits a homothetic Killing 
vector $\eta^A$ in $5D$. Namely\footnote{Here we use ${\cal{L}}_{\eta}g_{AB} 
= g_{A B,C}\eta^C + g_{C B}\eta^{C}_{,A} + g_{A C}\eta^{C}_{,B}$},
\begin{equation}
{\cal{L}}_{\eta}g_{AB} = 2 g_{AB},
\end{equation}
where
\begin{equation}
\eta^A = \frac{1}{2}(t,\;\;r,\;\;0,\;\;0,\;\;-\frac{3 + \alpha^2}{\alpha}y).
\end{equation}
Consequently, the   spacetime part of  (\ref{the solution}) inherits the 
so-called self-similar symmetry, where all dimensionless quantities in the 
solution can be put as functions of a {\em single} independent variable 
\cite{Cahill}-\cite{Jpdel5}. Specifically,
\begin{equation}
{\cal{L}}_{\xi}g_{\lambda \rho} = 2 g_{\lambda \rho},
\end{equation}
where
\begin{equation}
\xi^{\mu} = \frac{1}{2}(t,\;\;r,\;\;0,\;\;0).
\end{equation}
The self-similar nature of the spacetime metric becomes  evident in the 
canonical coordinates $\bar{t}$, $\bar{r}$  given by
\begin{equation}
t = \sqrt{A} B^{- b}{{\bar{t}}^{(1 - b)}}, \;\;\;r = \sqrt{A} B^{(b - 1)} {\bar{r}}^b, \;\;\;b \equiv 
\frac{1}{\sqrt{3 - \alpha^2}},
\end{equation}
where $A$ is some dimensionless constant.
Indeed, in these coordinates the $4D$ metric becomes
\begin{equation}
\label{4D solution}
ds^2 =  \frac{3 A^2(1 - b)^2}{\alpha^2}\xi^{- 2b}d{\bar{t}}^2  - \frac{A^2 }{(3 - \alpha^2)}{\xi}^{2(1 - 
b)}[d{\bar{r}}^2 + {\bar{r}}^2 d\Omega^2],
\end{equation}
where the similarity variable $\xi$ is
\begin{equation}
\xi = \left(\frac{\bar{t}}{\bar{r}}\right).
\end{equation}
It should be noted that the five-dimensional solution (\ref{the solution}) 
is {\em not} self-similar (in the sense mentioned above), only the $4D$ part 
is. Self-similar symmetry in general relativity generalizes the classical 
notion of similarity, which has successfully been applied in classical 
hydrodynamics to the description of shock waves created in explosions 
\cite{Sedov}, \cite{Barenblat}. The physical relevance of self-similar 
solutions is that they are singled out from a complicated set of initial 
conditions \cite{Cahill}.

\medskip

Secondly, the present solution differs from others in the literature in 
several ways:
\begin{enumerate}
\item Solution (\ref{the solution}) exists for both signs of the extra part 
of the metric, whereas most solutions that have drawn attention only satisfy 
$R_{AB} = 0$ for one sign or the other. For example, the standard $5D$ 
cosmology due to Ponce de Leon \cite{Jpdel6} only exists for signature $(+, 
-, -, -, -)$, while the wave solution due to Billyard and Wesson 
\cite{Billyard2} only exists for signature\footnote{For a recent 
discussion of the signature of the extra dimension see \cite{Jpdelsignature}.} $(+, -, -, -, +)$.

\item Many of the solutions in the literature are curved in $4D$ but 
Riemann-flat in $5D$ (see the catalog of Ref. $[6]$, pp. $95$ and $109$). 
However, metric (\ref{the solution}) is  {\em not} flat in $5D$. Indeed, 
$R_{ABCD}$ has $12$ nonzero components. 
\item The signs of the spacetime components of the Riemann tensor are the same for 
both signatures in (\ref{the solution}), whereas the signs of the components 
associated with the extra dimension reverse depending on the sign choice in 
(\ref{the solution}). 
\item The $5D$-Kretschmann scalar $K \equiv R_{ABCD}R^{ABCD}$ is unaffected by 
the signature. For (\ref{the solution}) we find $K \sim (tr)^{- 4}$, so it is singular on the hypersurface $rt = 0$. This kind 
of singularity is common is self-similar solutions. Physically this is 
because such solutions are applicable far from the origin and boundaries, 
where the initial conditions are unimportant \cite{Cahill}, \cite{Sedov}, 
\cite{Barenblat}.
\end{enumerate}

\section{Summary and conclusions}
The line element (\ref{the solution}) is an exact solution of the 
five-dimensional Einstein field equations in an empty $5D$ space. By 
Campbell's theorem it can be interpreted as a solution of the $4D$ field 
equations with energy-momentum (\ref{matter distribution}). Since the metric 
(\ref{the solution}) is independent of the extra coordinate, the trace of 
the energy-momentum tensor is zero and the matter distribution is 
radiation-like.

In the comoving frame the source can be modeled as an anisotropic fluid with 
heat flow in the radial direction. The parameter $\alpha$ in (\ref{matter 
distribution}) governs the evolution of the extra dimension and serves as a 
measure of the anisotropy. Namely, for $\alpha = \pm 1$ the fluid has 
isotropic pressure and the equation of state is that of radiation $\rho = 3 
p$, whereas for any other $\alpha$ the fluid has anisotropic pressures 
obeying $\rho = p + 2 p_{\perp}$.

An observer moving in the radial direction with velocity $v$ given by 
(\ref{radial velocity}) will interpret the matter distribution as an 
isotropic  fluid satisfying $\rho = 3p$,  with a radial heat flux directed outward, for any 
value of $\alpha$ in the regions $ 0 < \alpha \leq 1$ and $- \sqrt{3} < \alpha \leq - 1$. The energy density and the temperature are related by the Stefan-Boltzmann law (\ref{Stefan-Bolzmann law}). We note that  the introduction of energy flux 
is crucial here to account for $T_{01} \neq 0$. This is different from other 
solutions in the literature where $T_{01} \neq 0$ can be explained in terms 
of radial motion only.

The effective energy-momentum tensor satisfies the appropriate energy 
conditions but diverges at the hypersurfaces $rt = 0$. From a $5D$ 
viewpoint, the latter  is associated with the singularity of the Kretschmann scalar. But, from a $4D$ viewpoint it is 
related to the homothetic nature of the distribution, which is singled out 
from a complicated set of initial conditions.

An important feature here is that the matter density and pressure are 
spatially nonuniform. This suggest that our solution can be applied to 
describe the evolution of a spherical  inhomogeneity in an early universe.

We would like to  finish with the following remarks. Firstly, that although 
the $4D$-solution (\ref{4D solution}) was obtained in the context of STM in  
$5D$, by virtue of Campbell's theorem it is also a solution of the $4D$ 
Einstein field equations with energy-momentum (\ref{matter distribution}). 
Secondly, the same matter distribution is obtained if the metric (\ref{the 
solution}) is interpreted in the context of brane theory in $5D$ 
\cite{Jpdel3}. Finally, the metric (\ref{the solution}) exhibits a number of interesting properties, which make it different from other solutions in $5D$ gravity. We will discuss this in more detail elsewhere.   

\paragraph{Acknowledgments:} We would like to thank Kayll Lake for verifying 
the solution as well as for calculating the components of the Riemann tensor on the 
computer. This work was supported by the University of Puerto Rico (USA) and 
N.S.E.R.C. (Canada).

\end{document}